\documentclass[twocolumn,english,aps,prb,showpacs,floatfix,superscriptaddress]{revtex4}
\usepackage[T1]{fontenc}
\usepackage{amsmath,graphicx,hyperref}
\usepackage[latin9]{inputenc}

\begin{document}

\title{Relaxation and edge reconstruction in integer quantum Hall systems}

\author{Torsten Karzig}

\affiliation{\mbox{Dahlem Center for Complex Quantum Systems and
Fachbereich Physik, Freie Universit\"at Berlin, 14195 Berlin,
Germany}}

\author{Alex Levchenko}

\affiliation{Department of Physics and Astronomy, Michigan State University, East
Lansing, Michigan 48824, USA}

\author{Leonid I. Glazman}

\affiliation{Department of Physics, Yale University, 217 Prospect Street, New
Haven, Connecticut 06520, USA}

\author{Felix von Oppen}

\affiliation{\mbox{Dahlem Center for Complex Quantum Systems and
Fachbereich Physik, Freie Universit\"at Berlin, 14195 Berlin,
Germany}}

\date{\today}


\begin{abstract}
The interplay between the confinement potential and
electron-electron interactions causes reconstructions of Quantum
Hall edges. We study the consequences of this edge reconstruction
for the relaxation of hot electrons injected into integer quantum
Hall edge states. In translationally invariant edges, the relaxation
of hot electrons is governed by three-body collisions which are
sensitive to the electron dispersion and thus to reconstruction
effects. We show that the relaxation rates are significantly altered
in different reconstruction scenarios.
\end{abstract}
\maketitle

\section{Introduction}

The kinetic properties of one-dimensional (1D) quantum systems are
an active area of current research.\cite{review-1,review-2} What
makes the field exciting is that many-particle physics is
drastically different in one spatial dimension. This difference is
already evidenced in basic non-equilibrium properties such as the
microscopic mechanisms of relaxation. Within the scope of
Fermi-liquid theory, relaxation processes in higher dimensions
proceed by pair collisions of electrons which provide an efficient
mechanism for relaxation of initial nonequilibrium states. In
contrast, conservation of energy and momentum strongly restricts
scattering in one spatial dimension so that pair collisions
necessarily result in zero-momentum exchange or an interchange of
the momenta of the colliding particles. Neither process causes
relaxation. This poses the fundamental question of the microscopic
origin of relaxation in 1D systems. Notably, the absence of
relaxation by pair collisions, which holds regardless of the
strength of interaction, has received experimental support.
\cite{Barak}

The question of equilibration emerges in a diverse set of 1D
many-body systems. These include energy and momentum-resolved
tunneling experiments with nanoscale quantum
wires,\cite{Mason,Barak} quench dynamics of cold atomic
gases,\cite{Kinoshita,Hofferberth} as well as energy-spectroscopy
experiments on quantum Hall edge states driven out of
equilibrium.\cite{Altimiras10,Sueur10,Altimiras10b,Paradiso11} The
present paper is motivated by the latter experiments which are
carried out in a high-mobility two-dimensional electron system at
Landau level filling factor $\nu=2$. This system hosts two
co-propagating edge states which can be driven out of equilibrium by
inter-edge tunneling in the vicinity of quantum point contacts. This
generates a nonequilibrium distribution of electron energy (in the
sense of electronic edge transport) downstream from the contact
which is monitored as a function of the propagation distance by
means of a quantum-dot-based energy spectrometer. The experiments
show that the initial nonequilibrium distribution relaxes to a
stationary form which is close to the thermal distribution but with
an effective temperature and chemical potential.

Edge-state equilibration was also probed in experiments at
Landau-level filling factor $\nu=1$.\cite{Granger} Heat is carried
unidirectionally by the single chiral edge mode as confirmed by
thermopower measurements along the edge. These experiments found
that hot electrons injected locally into the edge cool down while
propagating along the edge. It is worth emphasizing that the
standard chiral-Luttinger-liquid model for quantum Hall edge
states\cite{Wen} does not account for equilibration effects. Indeed,
this model is exactly solvable and as usual, its integrability is an
obstacle to thermalization. In early works~\cite{Kane-Fisher} this
apparent difficulty was overcome by assuming a disordered edge where
impurity mediated scattering allows for interchannel equilibration.

These experimental discoveries led to a flurry of theoretical
activity. We briefly summarize these contributions and place our
work into their context. Two initial
publications\cite{Lunde10,Degiovanni10} used entirely different
concepts. Ref.\ \onlinecite{Lunde10} was based on a Boltzmann
kinetic equation for a disordered edge. Since
translation invariance is broken, momentum is no longer a good
quantum number and relaxation becomes possible even by two-particle
collisions. Ref.\ \onlinecite{Degiovanni10} adopted a bosonization
approach and combined it with a phenomenological model for the
plasmon distribution generated at the quantum point contact. Within
this model, thermalization was interpreted as a consequence of
plasmon dispersion which causes the electron wave packets to broaden
as they propagate with different group velocities. This picture was
elegantly elaborated and extended in Refs.\
\onlinecite{Kovrizhin11,Kovrizhin11b,Levkivskyi12}. A third
mechanism was proposed in the context of electronic Mach-Zehnder
interferometers~\cite{MZI-1,MZI-2} based on electron-plasmon
scattering. \cite{Heyl10} This mechanism relies on scattering of
high-energy electrons by low-energy plasmons enabled by the
curvature of the fermionic spectrum.

Despite the insight provided by these theories, important issues need to be
sorted out. First, these works do not give a definitive answer whether
relaxation is possible in translationally-invariant clean edges. Specifically,
the dispersion of plasmon modes may lead to a steady state but does not
constitute true relaxation as the energy in each plasmon mode is conserved.
Second, the edge of quantum Hall systems can be reconstructed due to Coulomb
interactions. The precise nature of reconstruction depends on the steepness of
the confinement potential, ranging from no reconstruction for very sharp
confinement potentials \cite{Halperin82} to alternating compressible and
incompressible stripes for very smooth edges.\cite{Chklovskii92} Indeed,
experiments\cite{Barak10,Deviatov11} point towards an important role of
reconstruction effects in energy transfer along the edge.

The purpose of the present study is to address these issues within
minimal models of unreconstructed and reconstructed edges.
Specifically, we consider energy relaxation of a hot particle
injected into translationally invariant quantum Hall edges at Landau
level filling factors  $\nu=1,2$. With the assumption that the
velocity $v_1$ of the injected particle differs sufficiently from
the Fermi velocity $v_F$, we treat the Coulomb interaction
perturbatively.\cite{Karzig10,Micklitz11} In this limit, relaxation
processes are dominated by three-body collisions which depend
sensitively on the electron dispersion and hence on the edge
reconstruction. We begin with a discussion of energy relaxation for
the unreconstructed edge in Sec.\ \ref{unreconstructed}. We then
discuss two simple models of reconstructed quantum Hall edges. In
Sec.\ \ref{spinreconstructed}, we discuss relaxation processes for a
spin-reconstructed edge for filling factor $\nu=2$. In Sec.\
\ref{chargereconstructed}, we turn to a minimal model of charge reconstruction
of a $\nu=1$ edge which provides the simplest realization of counter-propagating
edge modes. We conclude in Sec.\ \ref{conclusions}.

\section{Unreconstructed edge}
\label{unreconstructed}

\begin{figure}[t]
\begin{centering}
\includegraphics[scale=0.5]{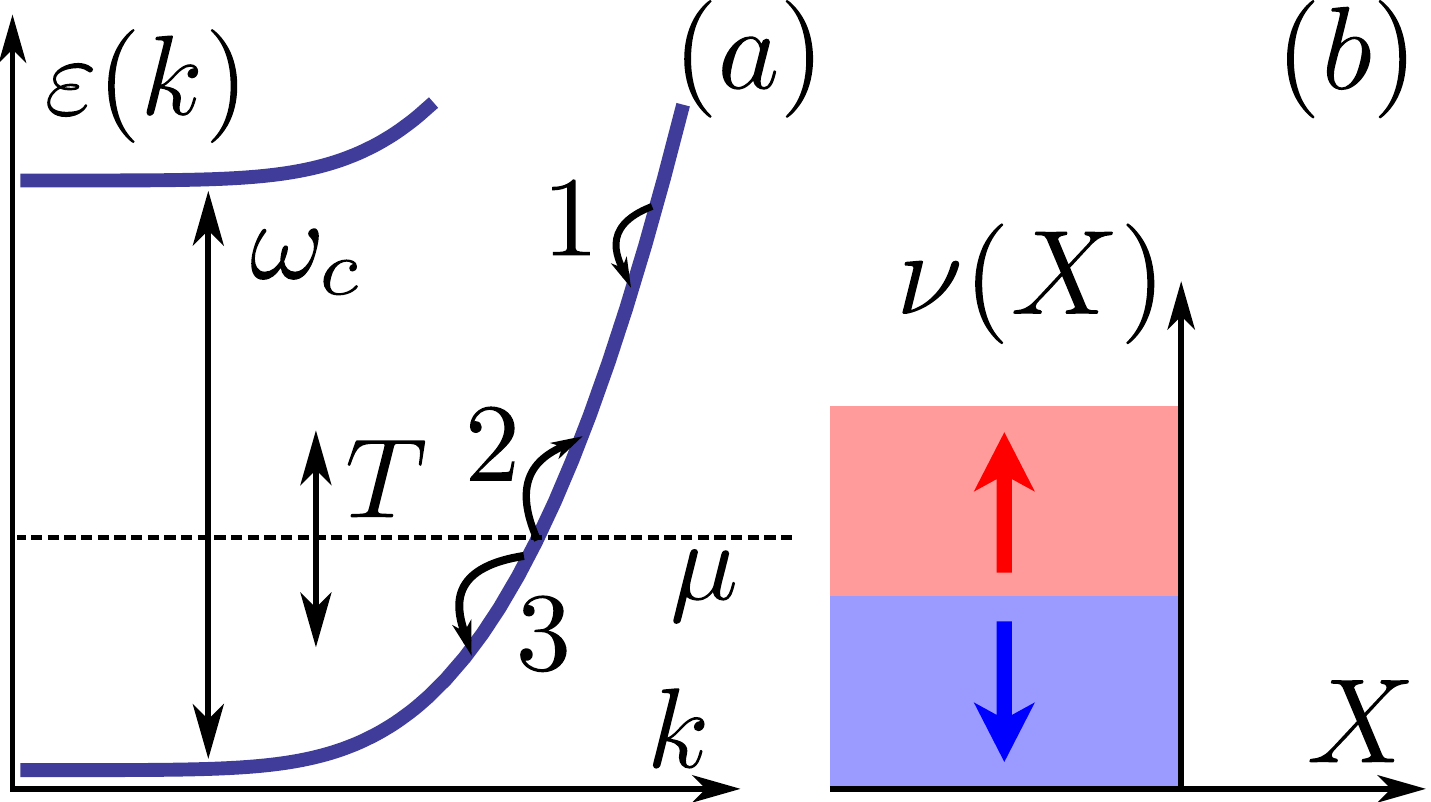}
\end{centering}
\caption{Edge without reconstruction effects. (a) Single particle
dispersions and typical relaxation processes in the the lowest
Landau level for a spin degenerate system. Vertical arrows label
temperature $T$ and cyclotron energy $\omega_c=eB/m$. (b) Shows the
corresponding sharp ($T=0$) occupation number in terms of the
guiding center position $X$.} \label{fig:sharp}
\end{figure}

A confinement potential $V_c(x)$ that is sharp on the scale of the Coulomb
interaction (i.e., $V_c^{'}\gg e^2/(\kappa l_B^2)$, where $\kappa$ is the
dielectric constant and $l_B$ denotes the magnetic length) remains stable
against interaction-induced reconstructions and the electron dispersion
$\varepsilon(k)$ can be obtained approximately from the noninteracting
Schr\"{o}dinger equation.\cite{Halperin82} A generic electronic dispersion of an
unreconstructed edge is sketched in Fig.\ \ref{fig:sharp}a, exhibiting a
confinement-induced bending of the Landau levels near the edge of the sample.

In the limit of high magnetic fields ($V_c^{'}\ll \omega_c/l_B$), the electron
states near the edge can be described by the lowest Landau level wave functions
\begin{equation}
 \psi_X(x,y)=(L l_B \sqrt{\pi})^{-1/2}e^{-i k y}e^{-(x-X)^2/2l_B^2},
\label{wavefunctions}
\end{equation}
in the Landau gauge. Here, $k=X/l_B^2$ and $L$ denotes the length of the sample
edge (taken along the $y$ direction). The defining feature of the
unreconstructed edge is the sharp zero-temperature occupation function
$\nu_\sigma(X)=\Theta(-X)$ of Landau level states with guiding center $X$ when
the Zeeman splitting $\varepsilon_Z$ is negligible [see Fig.\ \ref{fig:sharp}b].

The single particle dispersion near the Fermi energy (corresponding
to momentum $k_F$) is controlled by the confinement potential and
can be approximated as
\begin{equation}
 \varepsilon(k)=v_F(k-k_F)+(k-k_F)^2/2m_c\,.
 \label{sharp_dis}
\end{equation}
The dispersion is parametrized through the edge velocity $v_F=V_c^{'}l_B^2$ at
the Fermi energy and the curvature $1/m_c=V_c^{''}l_B^4$. Note that these
parameters become maximal for an infinitely sharp edge for which $v_F\sim
\omega_c l_B$ and $1/m_c\sim 1/m$.\cite{Halperin82} Note however, that a
description in terms of the wave functions in Eq.~(\ref{wavefunctions}) is no
longer valid in this extreme limit.

The finite curvature of the dispersion implies that at least three
particles are required for an energy and momentum conserving
relaxation process. Relaxation of a high-energy electron (labeled by
$i=1$ in Fig.\ \ref{fig:sharp}a) is possible by scattering two
electrons (labeled $i=2,3$ in Fig.\ \ref{fig:sharp}a) near the Fermi
energy. Indeed, due to the curvature of the dispersion near the
Fermi energy, exciting electron $i=2$ from the Fermi energy requires
more energy than scattering electron $i=3$ deeper into the Fermi
sea. Clearly, this relaxation process relies on finite temperature
and typical energy transfers for electrons $i=2,3$ at the Fermi
energy are of the order of $T$. Quantitatively, this process can
relax the hot particle with excess energy $\varepsilon\approx
v_F(k_1-k_F)=v_F\Delta k$ by $q_1=q_3 (k_{2'}-k_3)/\Delta k$, where
$q_i=k_{i'}-k_i$ is the momentum transferred to particle $i$ in the
collision. Note that $q_1\ll q_3$ so that relaxation occurs in many
small steps $v_F q_1\sim T^2/\varepsilon$.

For Landau-level filling factor $\nu=2$, these considerations apply when the
Zeeman splitting is small compared to temperature. In the opposite limit
$\varepsilon_Z\gg T$, the curvature of the dispersion implies that the Fermi
momenta and hence the Fermi velocities differ for the two spin directions. In
this case, relaxation is dominated by processes in which the electrons $i=1$ and
$i=2$ have opposite spins, and thus different Fermi momenta $k_{Fj}$ and Fermi
velocities $v_j$ with $j=1,2$. To include a finite Zeeman splitting at Landau
level filling factor $\nu=2$ as well as for later convenience, it is thus
beneficial to consider a modified dispersion
\begin{equation}
 \varepsilon(k)= \begin{cases} v_j(k-k_{Fj}), & k\approx k_{Fj}\\
 v_1(k-k_1)+\varepsilon, & k\approx k_1 \end{cases}
 \label{lin_dis}
\end{equation}
which is linearized in the vicinity of each of the three particles, including
the hot particle with velocity $v_1$ and momentum $k_1$. This captures the
behavior in the regime of strong Zeeman splitting $\varepsilon_Z\gg T$ on which
we will focus in the following. Nevertheless, we can also recover the results
for the quadratic dispersion and weak Zeeman splitting $\varepsilon_Z\ll T$ by
identifying $v_2-v_3$ with the typical velocity difference $T/(v_F m_c)$ due to
the curvature of the dispersion.

Using the dispersion in Eq.~(\ref{lin_dis}), energy and momentum conservation leads to
\begin{equation}
 q_1=\frac{v_2-v_3}{v_1-v_2}q_3\,.
\label{q1}
\end{equation}
The velocity difference $v_2-v_3= \varepsilon_Z/(v_2 m_c)$ is
controlled by the Zeeman splitting which we assume to be small
compared to the excitation energy $\varepsilon$ such that
$(v_2-v_3)\ll(v_1-v_2)$.

\begin{subsection}{Three-body scattering formalism}

Energy relaxation by processes of the kind shown in Fig.\ \ref{fig:sharp}a was
already discussed in the context of quantum wires in Ref.\
\onlinecite{Karzig10}. While our calculation here follows the same outline,
there are characteristic differences related to the nature of the interaction
matrix elements. The energy relaxation rate via three-body collisions is again
given by
\begin{equation}
 \frac{1}{\tau_E}=\sum_{231'2'3'}\frac{-v_1 q_1}{\varepsilon} W_{1'2'3'}^{123} n_2 n_3
(1-n_{1'})(1-n_{2'})(1-n_{3'})\,.
\label{energyrate}
\end{equation}
where $n_i$ is the Fermi-Dirac distribution function at $k_i$. The factor
involving $q_1$ weights the out-scattering rate with the relative relaxed
energy, accounting for the fact that the hot particle relaxes only a fraction of
its energy in a single collision. The three-body matrix element can be
evaluated by the generalized golden rule
\begin{equation}
W^{123}_{1'2'3'}=2\pi|\langle1'2'3'|VG_{0}V|123\rangle_{c}|^{2}\delta(E-E').
\end{equation}
Here, $G_0$ is the free Green's function,
$V=({1}/{2L})\sum_{k_{1}k_{2}q\sigma_{1}\sigma_{2}}V_{q}(k_1\!-\!k_2)a_{k_{1}+q\sigma_{1}}^{\dagger}
a_{k_{2}-q\sigma_{2}}^{\dagger}a_{k_{2}\sigma_{2}}a_{k_{1}\sigma_{1}}$
is the generic two-body interaction potential, and the subscript $c$
emphasizes that only connected processes contribute which involve
all three particles.

The calculation for quantum Hall edges differs from that for quantum wires in
the form of the Coulomb matrix element $V_q(k_1\!-\!k_2)$ which now has to be
evaluated using the Landau level wave functions in Eq.~(\ref{wavefunctions}).
For quantum Hall systems, the Coulomb matrix element is exponentially suppressed
by a factor of $\exp(-q^2l_B^2/2)$ for large momentum transfers. This is
especially relevant because large momentum transfers yield the leading
contribution to relaxation in quantum wires.\cite{Karzig10} Moreover,
$V_q(k_1\!-\!k_2)$ does not only depend on the momentum transfer but also on the
initial momentum difference which controls the distance between the guiding centers
of the interacting electrons. Focusing on the remaining low momentum transfer
processes ($q\ll 1/l_B$), one obtains (see Appendix-\ref{App-Vq} for details)
\begin{equation}\label{V}
V_q(k_1-k_2)\simeq\begin{cases} -\frac{2e^2}{\kappa}\ln|ql_B|,&\!\!k_1-k_2\ll l_B^{-1}\\
-\frac{2e^2}{\kappa}\ln|q(k_1-k_2+q)l_B^2|,&\!\!k_1-k_2\gg l_B^{-1}
\end{cases}
\end{equation}
with the understanding that at small $q$, the matrix elements will be eventually
cut off by a large length scale $\lambda \gg l_B$ which is given by the distance
to a screening gate. For $k_1-k_2\ll1/l_B$ the Coulomb matrix element is that of
a quantum wire of width $l_B$. For $k_1-k_2\gg 1/l_B$, the interaction is that
of electrons in two quantum wires separated by a distance of $(k_1-k_2+q)l_B^2$
which equals the average of the guiding center distances of the electrons before
and after the collision.

With the absence of large momentum transfer processes the three-body scattering
is dominated by the direct matrix element. The importance of the $(k_1\!-\!k_2)$
dependence of the Coulomb matrix element can be seen from the fact that the
linearized dispersion of Eq.~(\ref{lin_dis}) leads to a vanishing direct matrix
element for a quantum-wire-like Coulomb interaction $V_q(0)$ (see Appendix \ref{App-T}). In
contrast, when reiterating the derivation\cite{Lunde07} of the direct matrix
element $T^{123}_{1'2'3'}$ including the dependence of $V_q(k_1-k_2)$ on the initial momenta, the
result does {\em not} vanish and takes the form
\begin{eqnarray}
T^{123}_{1'2'3'}\!\!\! & = &\!\!\! -\frac{2
e^2}{L^{2}\kappa}\Bigg(\frac{k_{2}-k_{3}}{v_{2}-v_{3}}\frac{V_{q_{3}}(k_{2}-k_{3})}{(\Delta
k)^2}\label{VV2}\\
 &  & -\frac{2
V_{q_{3}}(\Delta k)}{\left(v_1-v_2\right)\left(\Delta k\right)}+\frac{v_{2}-v_{3}}{ k_ { 2 }
-k_{3}}\frac{V_{q_{1}}(\Delta k)}{(v_1-v_2)^2}\Bigg)\,.\nonumber
\end{eqnarray}
Here we used $k_1-k_2\approx k_1-k_{F2}=\Delta k$. This expression is applicable
under the assumption that all initial momentum-differences are large compared to
$1/l_B$ to also suit the reconstruction effects that will be discussed later.
For the unreconstructed edge, it is however more reasonable to assume
$k_2-k_3\ll1/l_B$ (for typical $\varepsilon_Z,T \ll e^2/\kappa l_B$) in which
case the last term of
Eq.~(\ref{VV2}), involving $V_{q_1}$, does not show up (see Appendix \ref{App-T}).

\end{subsection}

\begin{subsection}{Results for the unreconstructed edge}

For the unreconstructed edge, the momentum and velocity differences are linked
by the curvature of the confinement potential via $v_2-v_3=(k_2-k_3)/m_c$ and
$v_1-v_2=\Delta k/m_c$. The direct matrix element then takes the form
\begin{equation}
 T^{123}_{1'2'3'} = -\frac{2 e^2 m_c}{L^{2}\kappa (\Delta k)^2}\left[V_{q_{3}}(k_{2}-k_{3})-2V_{q_3}(\Delta k)\right]\,.
 \label{T}
\end{equation}
Since for large Zeeman energy the particles at $k_2$ and $k_3$ have opposite
spins, there is no exchange contribution (remember that exchange is appreciable
for small momentum transfers only) and Eq.~(\ref{T}) fully determines the three-body
matrix element. The corresponding energy relaxation rate can then be
obtained by power counting which yields
\begin{equation}
\frac{1}{\tau_E} \sim m_c(v_2-v_3)^2\left(\frac{e^2}{\kappa v_2}\right)^4
\frac{(m_c v_1^2)^4 T^4}{\varepsilon^8}\,,
\label{tauE}
\end{equation}
Here $1/m_c=V_c^{''}l_B^4$, $(v_2-v_3)=\varepsilon_Z/(v_2 m_c)$ and we also used
$\Delta k=m_c(v_1-v_2)=\varepsilon / v_1$.
In obtaining Eq.~\eqref{tauE}, a factor of $L/(v_1-v_2)$ emerges from eliminating the energy
$\delta$-function in Eq.~(\ref{energyrate}), each summation over the remaining
$k_2,k_3,q_3$ contributes a phase space factor of $\sim T/v_2$ and the weighting
factor $v_1 q_1/\varepsilon$ takes the form $v_1(v_2-v_3)T/[v_2(v_1-v_2)\varepsilon]$. Finally we
have to account for the competition between excitation ($q_1>0$) and relaxation
($q_1<0$) of the hot particle. The latter is slightly favored because the momentum
transfer working against the Fermi distribution is reduced by a fraction
$v_2 q_1/T\sim (v_2-v_3)/(v_1-v_2)$.

Equation~(\ref{tauE}), valid at $\varepsilon \gg m_c v_1 e^2/ \kappa$ and $\varepsilon_Z\gg T$ implies that the relaxation rate is strongly temperature dependent and can be enhanced by
increasing the magnetic field. 

As mentioned above, the relaxation rate in the opposite
limit of weak Zeeman splitting $\varepsilon_Z\ll T$ can be obtained up to
prefactors by replacing $(v_2-v_3) \sim T/(v_2 m_c)$. Note that this regime
allows for a low momentum transfer exchange term $T^{123}_{1'3'2'}$ because the
particles $2$ and $3$ are no longer necessarily of opposite spin.
$T^{123}_{1'3'2'}$ can then be obtained from Eq.~(\ref{T}) by replacing
$q_3\rightarrow k_{2'}-k_3$, which does not change the power counting argument.

It is therefore possible to combine both cases by setting $v_2-v_3=\mathrm{max}\{\varepsilon_Z,T\}/(v_2 m_c)$.
In the case $v_1\approx v_2=V_c^{'}l_B^2$ it is then possible to rewrite Eq.~\eqref{tauE} as
\begin{equation}
 \frac{1}{\tau_E} \sim \frac{\mathrm{max}\{\varepsilon_Z,T\}^2}{V_c^{''}l_B^2} \bigg(\frac{V_c^{'}}{V_c^{''}l_B}\bigg)^2\left(\frac{e^2}{\kappa l_B}\right)^4 \frac{T^4}{\varepsilon^8}\,,
 \label{tauE2}
\end{equation}
which applies in the regime $V_c^{'} l_B^2 \gg \varepsilon(V_c^{''}l_B^2/V_c^{'}) \gg e^2/\kappa$.

For the later comparison of the relaxation rates before and after edge reconstruction it will be useful
to consider the unreconstructed case as the $v_1\gg v_2\sim e^2/\kappa$ limit of Eq.~\eqref{tauE} [which can be applied for $\varepsilon \gg (e^2/\kappa l_B)^2/(V_c^{''}l_B^2)$]. Formally, this regime leaves the condition of applicability for the Taylor expansion of the confinement potential that defines $1/m_c=V_c^{''}\!(\mu)l_B^4$ and would lead to another inverse mass $V_c^{''}\!(\mu+\varepsilon)l_B^4$ for curvature effects at energies of the order of $\varepsilon$. Distinguishing these different masses does however not lead to qualitative changes of the results and for brevity of the presentation we assume a quadratic confinement potential over the energy interval $[\mu,\mu+\varepsilon]$. We can then rewrite Eq.~\eqref{tauE} as
\begin{equation}
 \frac{1}{\tau_E^{(u)}} \sim \left(\frac{\mathrm{max}\{\varepsilon_Z,T\}}{e^2/(\kappa l_B)}\right)^2 \left(V_c^{''}l_B^2\right)  \left(\frac{T}{\varepsilon}\right)^4\,,
 \label{tauEu}
\end{equation}
where we used that in this regime $m_c v_1^2\sim\epsilon$. The crossover between Eqs.~\eqref{tauE2} and \eqref{tauEu} can be obtained at their limits of applicability by setting $\varepsilon=e^2 V_c^{'}/(\kappa l_B^2 V_c^{''})$ and $V_c^{'}=e^2/(\kappa l_B^2)$. 

Note that for a spin polarized edge, Eqs.~(\ref{tauE})-(\ref{tauEu}) only apply if
the Coulomb interaction is not screened for momenta of the order of
$T/v_2$. For a screened short range interaction ($T/v_2\ll 1/\lambda$),
the Pauli principle then leads to a suppression of the energy
relaxation rate by an additional factor of $(T \lambda/v_2)^4\ll
1$.~\cite{Micklitz11}

\end{subsection}

\section{Spin reconstruction}
\label{spinreconstructed}

\begin{figure}[t]
\begin{centering}
\includegraphics[scale=0.5]{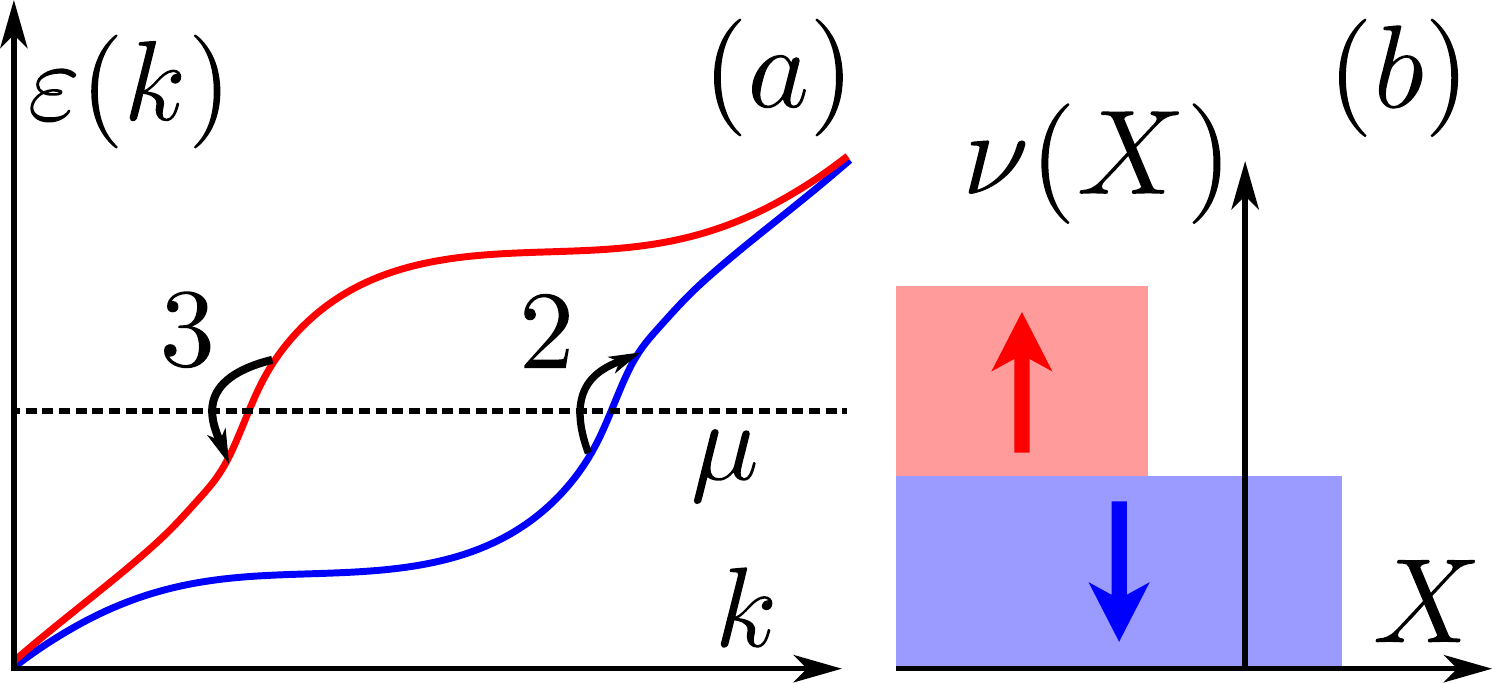}
\end{centering}
\caption{Spin reconstructed edge for $\varepsilon_Z=0$. (a) depicts the Hartree-Fock single
particle dispersion in the reconstructed region. Note that we set the curvature of the confinement
potential to zero such that the velocity difference $v_2-v_3>0$ is not obvious from the figure. (b)
shows the $T=0$ occupation numbers of the different spin species in terms of the guiding center
coordinate $X$.}
\label{fig:spin}
\end{figure}

Edge reconstruction in quantum Hall systems results from the competition between
the Coulomb interaction and the confinement potential. Spin reconstruction at
$\nu=2$ takes place when the confinement potential $V_c$ varies sufficiently
slowly so that $V_c^{'}<e^2/\kappa l_B^2$ and can
be understood at the level of the Hartree-Fock approximation.\cite{Dempsey93,Chamon94,Bauer95,Barlas11}
Once the slope of the confinement potential becomes weaker than
that of the repulsive Hartree potential $V_H$, it is favorable to deposit
charges outside the edge. This can be done without paying exchange energy by a
relative shift of the Fermi momenta of spin up and spin down particles, as
depicted in Fig.\ \ref{fig:spin}. In the absence of a Zeeman splitting,
$\varepsilon_Z=0$, this is a second order phase transition with spontaneous
breaking of the spin symmetry. Then, the distance of the two Fermi momenta
varies as $k_{F2}-k_{F3} \propto (|V_H^{'}|-V_c^{'})^{1/2}$, eventually saturating
at $\sim1/l_B$.\cite{Dempsey93} For finite Zeeman splitting $\varepsilon_Z$, the
spin symmetry is lifted by the Zeeman field and the transition is smeared on the
scale of $k_{F2}-k_{F3}\sim\varepsilon_Z/v_2$.

Spin reconstruction leads to characteristic changes in the single particle
dispersion that develops an "eye structure"[cf.\ Fig.\ \ref{fig:spin}a].
Important for the relaxation dynamics is the increase of
$v_2-v_3=(k_{F2}-k_{F3})/m_c$, which enhances the
typical energy transferred per step of relaxation [cf.~Eq.~(\ref{q1})].

For truly long range interactions, the Hartree-Fock approximation
predicts a logarithmic singularity $\sim e^2/\kappa
\ln(|k-k_F|l_B)$ of the particle velocity at the Fermi energy, which
is however cut off in the presence of screening, say by a nearby
gate electrode. The Fermi velocity is thus still of the order of
$v_2,v_3\sim e^2/\kappa$ for typical choices of the screening
length.

Even with spin reconstruction, the relaxation of hot particles can be
described within the model dispersion of Eq.~(\ref{lin_dis}). We consider the
case where the hot particle (not shown in Fig.\ \ref{fig:spin}) is injected well
outside the energy window $e^2/(\kappa l_B)$ of the reconstructed region.
This is compatible with the condition for the validity of a perturbative expansion,
which reduces to $v_1\gg v_2$ for the case of the Fermi velocity determined by the interaction.

The energy relaxation rate $1/\tau_E^{(s)}$ can now be derived in the same way
as for the unreconstructed edge and consequently, Eq.~(\ref{tauE}) also applies
to spin reconstructed edges. The crucial difference is that the velocity
difference $v_2-v_3$ is now strongly enhanced by the spin reconstruction, taking values
up to $v_2-v_3\sim 1/(m_c l_B)$. Comparing the rates before [$v_2-v_3\sim
\mathrm{max}\{\varepsilon_Z,T\}/(m_c v_2)$] and well after spin reconstruction, we find
an enhancement of the relaxation rates given by
\begin{equation}
\frac{1}{\tau_E^{(s)}} \sim \left(\frac{e^2/(\kappa
l_B)}{\mathrm{max}\{\varepsilon_Z,T\}}\right)^2 \frac{1}{\tau_E^{(u)}}.
\label{tau_spin}
\end{equation}

\section{Charge reconstruction}
\label{chargereconstructed}

\begin{figure}[t]
\begin{centering}
\includegraphics[scale=0.5]{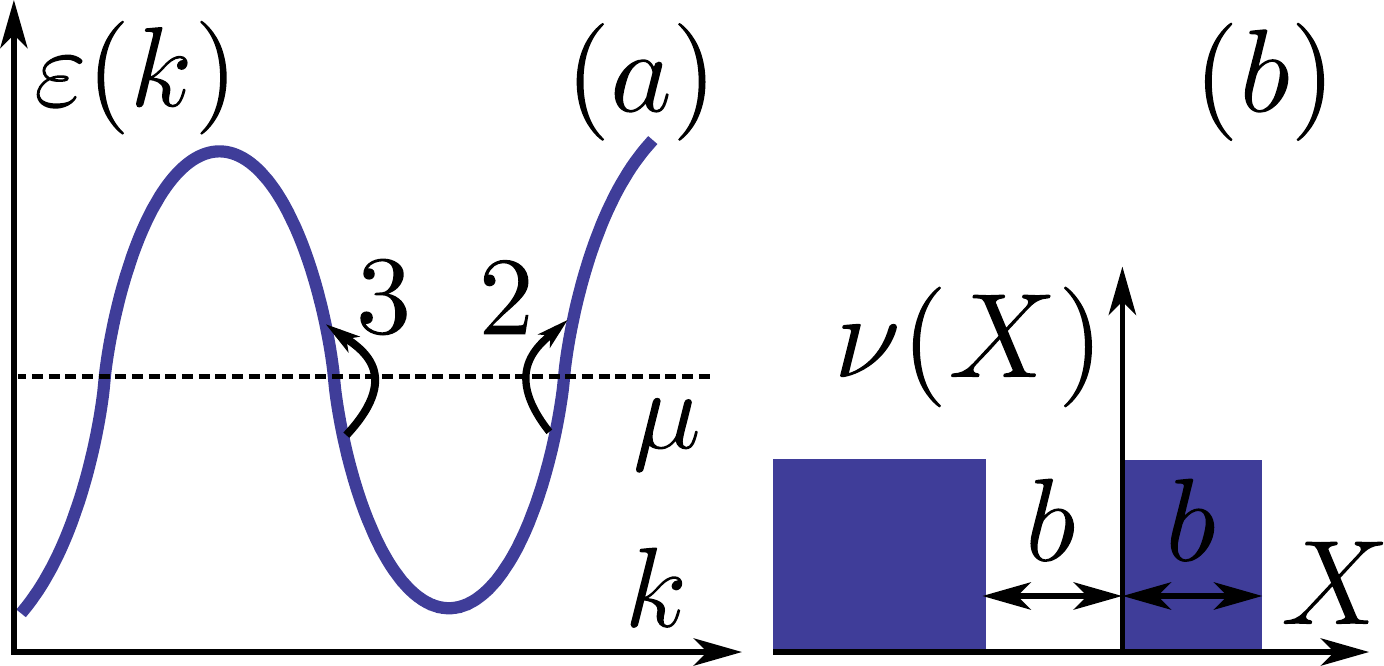}
\end{centering}
\caption{Charge reconstructed edge. (a) depicts the Hartree-Fock single particle
dispersion in the reconstructed region of a spin polarized sample at $\nu=1$. (b)
shows the occupation number near the edge of the sample.}
\label{fig:charge}
\end{figure}

For confinement potentials that vary even more smoothly, changing by
$e^2/\kappa l_B$ over a region $w>l_B$, charge reconstruction may occur
such that part of the electrons at the edge are pushed away from the bulk by a length
of the order of $l_B$. \cite{Chamon94,Barlas11} It leads to
a non-monotonic behavior of the dispersion with momentum and the
creation of two additional counter-propagating\cite{conductance}
edge modes, as depicted in Fig.\ \ref{fig:charge}.

A minimal model for charge reconstruction considers filling factor $\nu=1$
within the Hartree-Fock approximation.\cite{Chamon94} It is convenient to formally
model the confinement potential by a positive background charge which is distributed
spatially as if it was occupying lowest Landau level wave functions $\psi_X$ with occupation
numbers $\nu_c(X)=\Theta(-X)$. The advantage of this model is that such a confinement
potential exactly cancels the Hartree potential of the electrons for an unreconstructed
edge. In this case, the electron occupation of the unreconstructed edge is stabilized by the
(attractive) exchange potential.

The reconstruction transition can then be modeled by changing the
abrupt drop of $\nu_c(X)$ into a linear decrease over a length $w$.
For the unreconstructed electron occupations, this leads to negative
(at $X<0$) and positive (at $X>0$) excess charges, causing a dipole
field that favors separating electrons from the bulk. Once this
dipole field overcomes the exchange potential, a charge reconstruction
transition takes place. Within the Hartree-Fock approximation, this
happens for $w\sim8l_B$. Due to the particle-hole symmetric choice
of the confinement potential around $X=0$, the width and the
distance of the additional stripe from the bulk electron droplet
both take the same value $b$. Moreover, the transition is of first
order in the sense that $b$ changes abruptly at the transition from
zero to a value of the order of $l_B$.

Note that the same mechanism induces new (weaker) effective dipole
fields at each of the three Fermi points as the edge becomes yet
smoother. Thus, increasing $w$ even further causes additional
stripes to appear, eventually approaching the limit of a
compressible edge which is expected for $w\gg
l_B$.\cite{Chklovskii92} In the following we will focus on $w\gtrsim
l_B$, remaining well outside the compressible limit.

Energy relaxation in the charge reconstructed case can also be captured by the
dispersion (\ref{lin_dis}) when setting $v_3<0$ and choosing the particle $i=2$ to lie
in one of the co-propagating branches.\cite{branches} The three Fermi velocities of
the charge reconstructed edge are essentially determined by the variation of the
exchange potential, which is short ranged such that
$b\gtrsim l_B$ already approximates the bulk edge ($b\rightarrow \infty$)
behavior. Consequently, the magnitudes of the Fermi velocities are equal to that
of the unreconstructed edge and $\sim e^2/\kappa$. In line with the
discussions above, we consider the relaxation of a hot particle injected well
outside the reconstructed region with $v_1\gg e^2/\kappa$.

The nonmonotonic behavior of the dispersion introduces a new relaxation process
which relaxes the hot particle by exciting two counter-propagating electron-hole
pairs [see Fig.~\ref{fig:charge}]. This eliminates the restriction that the
energy transfers at the Fermi energy cannot exceed the temperature and makes
the relaxation process similar to that for non-chiral quantum wires.\cite{Karzig10} Unlike
for quantum wires, however, the momentum transfers at the Fermi energy of the
co- and counter-propagating branch are of the same order.

The three-body matrix element of Eq.\ (\ref{VV2}) still applies in the presence
of charge reconstruction because its derivation did not require a specific sign
of $v_3$. Note however that for the charge reconstructed case $v_2-v_3\sim v_2 \sim e^2/\kappa$ and is therefore not connected to $k_{F2}-k_{F3}\sim 1/l_B$ by the curvature of the confinement potential.  Assuming that there are no substantial curvature effects on the scale of the reconstructed region $V_c^{''}l_B^2 \ll e^2/(\kappa l_B)$ the last term of Eq.~\eqref{VV2} dominates the three-body matrix element and Eq.~\eqref{T} modifies to
\begin{equation}
 T^{123}_{1'2'3'} = -\frac{2 e^2}{L^{2}\kappa (\Delta k)^2}\frac{e^2/(\kappa l_B)}{(V_c^{''}l_B^2)^2 l_B^2} V_{q_{1}}(\Delta k)\,.
\end{equation}

The crucial difference for the energy relaxation rates compared to the unreconstructed case 
arises from the large allowed momentum $q_3\sim 1/l_B$, which is
limited only by the size of the reconstructed region for which the linearized
dispersion applies. This increases both the momentum phase space to $(L/l_B)^3$
and the typical relaxed momentum to $(v_2-v_3)/[(v_1-v_2) l_B]$. Moreover,
excitation and relaxation processes no longer need to be balanced when
$e^2/\kappa l_B\gg T$, and we find
\begin{equation}
\frac{1}{\tau_E^{(c)}} \sim \bigg(\frac{e^2/(\kappa l_B)}{V_c^{''}l_B^2}\bigg)^3 \sqrt{\varepsilon V_c^{''}l_B^2}\left(\frac{e^2/(\kappa l_B)}{\varepsilon}\right)^4
\label{tau_charge}
\end{equation}
which applies for $\varepsilon \gg (e^2/\kappa l_B)^2/(V^{''}l_B^2)$ and allows for relaxation even at $T=0$.   Equation~(\ref{tau_charge})
implies that the increased phase space and the energy relaxation step size leads to a dramatic enhancement of the relaxation rate compared to the unreconstructed case [see Eq.~(\ref{tauEu})] as
\begin{equation}
\frac{1}{\tau_E^{(c)}} \sim \sqrt{\frac{\varepsilon}{V^{''}l_B^2}}\bigg(\frac{e^2/(\kappa l_B)}{V_c^{''}l_B^2}\bigg)^3\left(\frac{e^2/(\kappa l_B)}{T}\right)^6 \frac{1}{\tau_E^{(u)}}\,,
\label{charge_compare}
\end{equation}
where we used the limit $T\gg \varepsilon_Z$.

\section{Conclusions}
\label{conclusions}

We studied three-body processes as an intrinsic mechanism for relaxation of hot
electrons in clean integer quantum Hall edges at Landau level filling factors
$\nu=1$ and $\nu=2$. These processes rely crucially on the form of the electron
dispersion and are thus susceptible to edge reconstruction effects. For an
unreconstructed edge, energy relaxation requires a finite temperature which
determines the phase space for the relaxation processes. The energy given up by
the hot electron in a single three-body collision is controlled by curvature
effects on the scale of temperature or Zeeman energy so that the relaxation rate
can be tuned by a magnetic field once $\varepsilon_Z\gg T$.

While unreconstructed edges are expected for steep confinement
potentials, smoother confinement potentials with $V_c^{'}\lesssim
e^2/(\kappa l_B^2)$ may lead to an interaction-induced spin
reconstruction, which causes a relative shift of the Fermi momenta
of the two spin species by $\sim 1/l_B$. The three-body processes
are then controlled by curvature effects on the scale of the
interaction energy $e^2/(\kappa l_B)$ which causes a strong
increase of the relaxation rate [see Eq.~(\ref{tau_spin})].

Even softer confinement may cause charge reconstruction which introduces
additional co- and counter-propagating edge modes. The presence of
counter-propagating modes allows for relaxation even at $T=0$. Consequently,
the phase space for three-body collisions is no longer controlled by temperature but
by the size of the reconstructed region $\sim e^2/(\kappa l_B)$ which ensues
an additional dramatic enhancement of the relaxation rate [see Eq.~(\ref{charge_compare})].

Experimental studies of interaction-induced reconstruction transitions in high
magnetic fields have been performed.\cite{Barak10} Our study suggests that it
would be rewarding to experimentally investigate relaxation processes in such
systems.

\begin{acknowledgments}
We acknowledge financial support of the Deutsche
Forschungsgemeinschaft under SPP 1538  (FUB) and by DOE under
Contract No. DEFG02-08ER46482 (Yale). A.~L. acknowledges support
from Michigan State University.
\end{acknowledgments}
\appendix

\section{Calculation of the Coulomb matrix element $V_q(k_1-k_2)$}\label{App-Vq}

Within this section we provide all essential details needed for the
derivation of Eq.~\eqref{V} presented in the main text. We assume
that the edge is smooth enough that we can approximate the electron
wave functions by those of the bulk. We start from the interaction
matrix element in real space
\begin{equation}
V_{dX}(X,X')=\left\langle\psi^{(1)}_{X+dX}
,\psi^{(2)}_{X'-dX}\left|\frac{e^2/\kappa}{|\mathbf{r}^{(1)}-\mathbf{r}^{(2)}|}
\right|\psi^{(1)}_{X},\psi^{(2)}_{X'}\right\rangle.
\end{equation}
In the following we will measure all lengths scales in units of
magnetic length $l_B$. In this units the guiding center coordinate
directly translates to momenta. With the lowest Landau level wave
functions of Eq.~\eqref{wavefunctions} we then find
\begin{eqnarray}
V_{dX}(X,X')=\frac{e^2}{\pi\kappa L^2}\int dxdyd\Delta xd\Delta y
\frac{e^{-\sqrt{\Delta x^2+\Delta y^2}/\lambda}}{\sqrt{\Delta
x^2+\Delta y^2}}\nonumber \\
e^{-i\Delta
ydX}e^{-\frac{1}{2}(x-X)^2}e^{-\frac{1}{2}(x-X-dX)^2}\nonumber\\
e^{-\frac{1}{2}(x+\Delta x-X')^2}e^{-\frac{1}{2}(x+\Delta
x-X'+dX)^2}
\end{eqnarray}
where we used the screened Coulomb potential which carries an extra
factor $e^{-\sqrt{\Delta x^2+\Delta y^2}/\lambda}$ with $\lambda$
being the distance to a screening gate. The integration over $\Delta
y$ gives $2K_0(|\Delta xdX|)$ in the case when $dX\gg1/\lambda$, where
$K_0$ is the Bessel function of imaginary argument. If, however,
$dX\ll1/\lambda$ the integral is cut off and the result changes to
$2K_0(|\Delta x/\lambda|)$. We will derive results for the
$dX\gg1/\lambda$ case and keep in mind appropriate changes for the
other limit. After $y$ integration, that gives a factor of $L$ we
obtain the intermediate step
\begin{eqnarray}
V_{dX}(X,X')=\frac{2e^2}{\pi\kappa L}e^{-\frac{1}{2}dX^2}\int
dxd\Delta x
K_0(|\Delta xdX|)\nonumber\\
e^{-\frac{1}{2}(2x-X-X'+\Delta x)^2}e^{-\frac{1}{2}(X-X'+dX+\Delta
x)^2}.
\end{eqnarray}
Performing now the Gaussian integral over $x$, that gives a factor of
$\sqrt{\pi/2}$, followed by using the Landau gauge to replace guiding
center coordinates by momenta one arrives at
\begin{equation}\label{Vqk}
V_q(k)=\sqrt{\frac{2}{\pi}}\frac{e^2}{\kappa}e^{-q^2/2}\int
d\xi e^{-\frac{1}{2}(k+q+\xi)^2}K_0(|\xi q|)
\end{equation}
where we used short-hand notation $X\!-\!X'=kl^2_B=k$ (with $l_B=1$). Note
that $V_q(k)=V_q(-k-2q)$ and is therefore not symmetric, which plays
an important role. We see immediately that scattering processes with
large momentum transfer $q\gg1$ are exponentially suppressed. We
therefore concentrate on the opposite limit of $q\ll 1$ when the
exponential prefactor $e^{-q^2/2}$ can be set to unity.

Let us study limiting cases of Eq.~\eqref{Vqk}. In the case when
$k\gg1$ one can approximate the exponential under the integral by the
delta-function $\sqrt{2\pi}\delta(k+q+\xi)$, and thus obtains
\begin{equation}
V_q(k)=\frac{2e^2}{\kappa}K_0(|q(k+q)|).
\end{equation}
Using the asymptotic form of the Bessel function and restoring units
of $l_B$ one recovers the second limit in Eq.~\eqref{V}.

In the other limiting case when $k\ll1$ one can approximate the exponential under the
integral of Eq.~\eqref{Vqk} by $e^{-\xi^2/2}$ and then complete
integration exactly with the result
\begin{equation}
V_q(k)=\frac{e^2}{\kappa}K_0(q^2/4).
\end{equation}
With the logarithmic accuracy at small $q$ this translates into the
first limit of Eq.~\eqref{V}.

\section{Calculation of the three-body matrix element $T^{123}_{1'2'3'}$}\label{App-T}

In general the three-particle scattering amplitude
$\langle1'2'3'|VG_{0}V|123\rangle_{c}$ contains six terms: one
direct and five exchange contributions.~\cite{Lunde07} As explained
in the text we need only the former one which reads
explicitly~\cite{Lunde07}
\begin{widetext}
\begin{eqnarray}
\hskip-.85cm
&&T^{123}_{1'2'3'}=\frac{\delta_{\Sigma,\Sigma'}}{L^2}\nonumber\\
\hskip-.85cm
&&\left[\frac{V_{k_{3'}-k_3}(k_3-k_2)V_{k_{1'}-k_1}(k_{2'}-k_{1'})}
{\varepsilon_{k_3}+\varepsilon_{k_2}-\varepsilon_{k_{3'}}-\varepsilon_{k_2+k_3-k_{3'}}}
+\frac{V_{k_{1'}-k_1}(k_1-k_3)V_{k_{2'}-k_2}(k_{3'}-k_{2'})}
{\varepsilon_{k_1}+\varepsilon_{k_3}-\varepsilon_{k_{1'}}-\varepsilon_{k_3+k_1-k_{1'}}}
+\frac{V_{k_{2'}-k_2}(k_2-k_1)V_{k_{3'}-k_3}(k_{1'}-k_{3'})}
{\varepsilon_{k_2}+\varepsilon_{k_1}-\varepsilon_{k_{2'}}-\varepsilon_{k_1+k_2-k_{2'}}}\right.
\nonumber\\
\hskip-.85cm
&&\left.+\frac{V_{k_{2'}-k_2}(k_2-k_3)V_{k_{1'}-k_1}(k_{3'}-k_{1'})}
{\varepsilon_{k_2}+\varepsilon_{k_3}-\varepsilon_{k_{2'}}-\varepsilon_{k_3+k_2-k_{2'}}}
+\frac{V_{k_{1'}-k_1}(k_1-k_2)V_{k_{3'}-k_3}(k_{2'}-k_{3'})}
{\varepsilon_{k_1}+\varepsilon_{k_2}-\varepsilon_{k_{1'}}-\varepsilon_{k_{2}+k_{1}-k_{1'}}}
+\frac{V_{k_{3'}-k_3}(k_3-k_1)V_{k_{2'}-k_2}(k_{1'}-k_{2'})}
{\varepsilon_{k_3}+\varepsilon_{k_1}-\varepsilon_{k_{3'}}-\varepsilon_{k_1+k_3-k_{3'}}}\right]
\end{eqnarray}
where the spin structure is
$\delta_{\Sigma,\Sigma'}=\delta_{\sigma_1,\sigma_{1'}}
\delta_{\sigma_2,\sigma_{2'}}\delta_{\sigma_3,\sigma_{3'}}$ and
the Coulomb matrix element $V_q(k)$ was derived in the preceding
section. Now using the dispersion relation from Eq.~\eqref{lin_dis},
and constrain on momentum transfers from Eq.~\eqref{q1}, imposed by the conservation
laws, one can simplify
$T^{123}_{1'2'3'}$ to
\begin{eqnarray}
T^{123}_{1'2'3'}\approx\frac{\delta_{\Sigma,\Sigma'}}{L^2}\left[\frac{V_{q_3}(k_3-k_2)V_{q_1}(k_1-k_2+q_3)-
V_{q_1}(k_1-k_2)V_{q_3}(k_3-k_2+q_1)}{q_3(v_2-v_3)}\right.+\nonumber\\
\frac{V_{q_2}(k_2-k_3)V_{q_1}(k_1-k_3+q_2)-V_{q_1}(k_1-k_3)V_{q_2}(k_2-k_3+q_1)}{q_2(v_3-v_2)}+
\nonumber\\
\left.\frac{V_{q_3}(k_3-k_1)V_{q_2}(k_2-k_1+q_3)-V_{q_2}(k_2-k_1)V_{q_3}(k_3-k_1+q_2)}{q_3(v_1-v_3)}\right]\,,
\end{eqnarray}
where we used the property $V_q(k)=V_q(-k-2q)$. It is important to stress
that the above expression would vanish by ignoring the dependence of the Coulomb
matrix element on initial momenta, namely for $V_q(k)=V_q$. To
proceed further we make use of the assumption that injected
particle is of high energy, such that $v_1\gg v_{2,3}$ and $k_1\gg
k_{2,3}$. In this case we expand $V_{q_i}(\Delta k+q_i)$ in $q_i$.
For the interaction $V_q(k)=-\frac{2e^2}{\kappa}\ln(|kq|l^2_B)$ we
obtain after the expansion
\begin{eqnarray}
T^{123}_{1'2'3'}\approx\frac{2e^2\delta_{\Sigma,\Sigma'}}{\kappa
L^2}\left[-\frac{V_{q_3}(k_3-k_2)}{(v_2-v_3)(k_1-k_2)}-\frac{V_{q_1}(k_1-k_2)}{(v_1-v_2)(k_2-k_3)}
+\frac{V_{q_2}(k_2-k_3)}{(v_2-v_3)(k_1-k_3)}\right.\nonumber\\
\left.+\frac{V_{q_1}(k_1-k_3)}{(v_1-v_3)(k_2-k_3)}+\frac{V_{q_3}(k_3-k_1)}{(v_1-v_3)(k_1-k_2)}
+\frac{V_{q_2}(k_2-k_1)}{(v_1-v_2)(k_1-k_3)}\right]\,.
\end{eqnarray}
\end{widetext}
Note that if we are in the regime when $k_2-k_3\ll l^{-1}_B$ we have
to use the interaction potential
$V_q(k)=-\frac{2e^2}{\kappa}\ln(|q|l_B)$, which has a vanishing
derivative with respect to $k$. This can be accounted for by
removing the two terms with $V_{q_1}(\ldots)$ in the above formula
for $T^{123}_{1'2'3'}$. Finally, to leading logarithmic order we
can set $V_{q_1}(k_1-k_2)=V_{q_1}(k_1-k_3)$ as well as
$V_{q_2}(k_2-k_1)=V_{q_3}(k_3-k_1)=V_{q_3}(k_1-k_2)$ and
$V_{q_3}(k_3-k_2)=V_{q_2}(k_2-k_3)$ to obtain Eq.~\eqref{VV2} since
the spin summation is equal to unity.


\begin{thebibliography}{References}
\bibitem{review-1}
V.~V.~Deshpande,  M.~Bockrath,  L.~I.~Glazman, and A.~Yacoby, Nature
\textbf{464}, 209 (2010).

\bibitem{review-2}
A.~Imambekov, T.~L.~Schmidt, L.~I.~Glazman, preprint arXiv:1110.1374
(to appear in Rev. Mod. Phys.).

\bibitem{Barak}
G.~Barak,  H.~Steinberg,  L.~N.~Pfeiffer,  K.~W.~West, L.~Glazman,
F.~von Oppen, and A.~Yacoby, Nature Physics \textbf{6}, 489 (2010).

\bibitem{Mason}
Y.-F.~Chen, T.~Dirks, G.~Al-Zoubi, N.~O.~Birge, and N.~Mason, Phys.
Rev. Lett. \textbf{102}, 036804 (2009).

\bibitem{Kinoshita}
T.~Kinoshita, T.~Wenger, and D.~S.~Weiss, Nature \textbf{440}, 900
(2006).

\bibitem{Hofferberth}
S.~Hofferberth, I.~Lesanovsky, B.~Fischer, T.~Schumm, and
J.~Schmiedmayer, Nature \textbf{449}, 324 (2007).

\bibitem{Altimiras10}
C.~Altimiras, H.~le~Sueur, U.~Gennser, A.~Cavanna, D.~Mailly, and
F.~Pierre, Nat. Phys. \textbf{6}, 34 (2010).
%experiment:Non-equilibrium edge-channel spectroscopy in the integer quantum Hall regime

\bibitem{Sueur10}
H.~le~Sueur, C.~Altimiras, U.~Gennser, A.~Cavanna, D.~Mailly, and
F.~Pierre, Phys. Rev. Lett. \textbf{105}, 056803 (2010).
%experiment:Energy Relaxation in the Integer Quantum Hall Regime

\bibitem{Altimiras10b}
C.~Altimiras, H.~le~Sueur, U.~Gennser, A.~Cavanna, D.~Mailly, and
F.~Pierre, Phys. Rev. Lett. \textbf{105}, 226804 (2010).
%experiment:Tuning Energy Relaxation along Quantum Hall Channels

\bibitem{Paradiso11}
N.~Paradiso, S.~Heun, S.~Roddaro, L.~Sorba, F.~Beltram, G.~Biasiol,
Phys. Rev. B \textbf{84}, 235318 (2011).

\bibitem{Granger}
G.~Granger, J.~P.~Eisenstein, and J.~L.~Reno Phys. Rev. Lett.
\textbf{102}, 086803 (2009).

\bibitem{Wen}
X.~G.~Wen, Phys. Rev. Lett. \textbf{64}, 2206 (1990); Phys. Rev. B
\textbf{43}, 11025 (1991).

\bibitem{Kane-Fisher}
C.~L.~Kane and M.~P.~A.~Fisher, Phys. Rev. B \textbf{51}, 13449
(1994); Phys. Rev. B \textbf{52}, 17393 (1995); Phys. Rev. B
\textbf{55}, 15832 (1997).

\bibitem{Lunde10}
A.~M.~Lunde, S.\ E.\ Nigg, and M.\ B\"{u}ttiker, Phys. Rev. B
\textbf{81}, 041311 (2010).
%disorder induced relaxation

\bibitem{Degiovanni10}
P.~Degiovanni, C.\ Grenier, G.\ F\`{e}ve, C.\ Altimiras, H.\ le\ Sueur,
and F.\ Pierre, Phys. Rev. B \textbf{81}, 121302 (2010).

\bibitem{Kovrizhin11}
D.~L.\ Kovrizhin and J.\ T.\ Chalker, Phys. Rev. B \textbf{84}, 085105
(2011).
%'relaxation' but no thermalization due to the plasmon dispersion or different plasmon velocities of the symmetric and asymmetric plasmon branch in a \nu=2 scenario.

\bibitem{Kovrizhin11b}
D.\ L.\ Kovrizhin and J.\ T.\ Chalker, arXiv:1111.3914 (2011).
%same idea as above but applied specifically to the Sueur10 experiment.

\bibitem{Levkivskyi12}
I.\ P.\ Levkivskyi and E.\ V.\ Sukhorukov, Phys. Rev. B \textbf{85},
075309 (2012).
%non-equilibrium bosozination for relaxation in QH edges. Also based on different velocities of the charged and dipole mode. No real thermalization.

\bibitem{MZI-1}
Y.~Ji, Y.~Chung, D.~Sprinzak, M.~Heiblum, D.~Mahalu, and
H.~Shtrikman, Nature \textbf{422}, 415 (2003).

\bibitem{MZI-2}
I.~Neder, M.~Heiblum, Y.~Levinson, D.~Mahalu, and V.~Umansky, Phys.
Rev. Lett. \textbf{96}, 016804 (2006).

\bibitem{Heyl10}
M.\ Heyl, S.\ Kehrein, F.\ Marquardt, and C.\ Neuenhahn, Phys. Rev. B
\textbf{82}, 033409 (2010).

\bibitem{Halperin82}
B.\ I.\ Halperin, Phys. Rev. B \textbf{25}, 2185 (1982).
%infinitely sharp edge

\bibitem{Chklovskii92}
D.\ B.\ Chklovskii, B.\ I.\ Shklovskii, and L.\ I.\ Glazman, Phys. Rev. B
\textbf{46}, 4026 (1992).
%electrostatics of smooth edge

\bibitem{Barak10}
G.\ Barak, L.\ N.\ Pfeiffer, K.\ W.\ West, B.\ I.\ Halperin, and A.\ Yacoby,
preprint arXiv:1012.1845.
%experiment:Spin reconstruction in quantum wires subject to a perpendicular magnetic field; they see signatures of the Dempsey reconstruction.

\bibitem{Deviatov11}
E.\ V.\ Deviatov, A.\ Lorke, G.\ Biasiol, L.\ Sorba, Phys. Rev. Lett.
\textbf{106}, 256802 (2011).

\bibitem{Karzig10} T.\ Karzig, L.\ I.\ Glazman, and F.\ von Oppen, Phys. Rev. Lett. {\bf 105}, 226407 (2010).
%Energy Relaxation and Thermalization of Hot Electrons in Quantum Wires

\bibitem{Micklitz11} T.\ Micklitz and A.\ Levchenko, Phys. Rev. Lett. {\bf 106}, 196402 (2011).
%Thermalization of Nonequilibrium Electrons in Quantum Wires

\bibitem{Lunde07} A.~Lunde, K.~Flensberg, and L.~I.~Glazman, Phys. Rev. B {\bf 75}, 245418 (2007).

\bibitem{Dempsey93}J.\ Dempsey, B.\ Y.\ Gelfand, and B.\ I.\ Halperin, Phys. Rev. Lett. \textbf{70}, 3639 (1993).
%edge reconstruction due to Hatree-Fock

\bibitem{Bauer95}T.\ H.\ Stoof and G.\ E.\ W.\ Bauer, Phys. Rev. B \textbf{52}, 12143 (1995).
%DFT calculation, more or less confirming Dempsey et al

\bibitem{Chamon94}C.\ de\ C.\ Chamon and X.\ G.\ Wen, Phys. Rev. B \textbf{49}, 8227 (1994).

\bibitem{Barlas11}Y.\ Barlas, Y.\ N.\ Joglekar, and K.\ Yang, Phys. Rev. B \textbf{83}, 205307 (2011).
%theoretical explaination of the barak10 experiment with Hartree Fock approximation. Instead of Dempsey, the assume a triangular confining potential in a regime where the Hall bar is better described as a wire. They find nice (qualitative) agreement with the experiment and also predict the formation of additional edge states (charge reconstruction) ala Chamon94 for smoother potentials.

\bibitem{conductance} Note that the conductance may stay at the $\nu=1$ level due to impurity induced elastic backscattering of electrons at the Fermi level. At higher energies inelastic scattering leads to a shorter mean free path and we can therefore omit the disorder-induced backscattering.

\bibitem{branches} Processes where the particle $i=2$ is on the inner or outer co-propagating branch contribute at the same order in the power counting argument. The choice depicted in Fig.~\ref{fig:charge} is slightly favored because of a larger Coulomb matrix element due to the smaller distance to the hot particle.

\end{thebibliography}
\end{document}